\documentclass[letterpaper,twocolumn,10pt]{article}
\usepackage{usenix2019,epsfig}
\usepackage[hyphens]{url}
\usepackage{hyperref}
\usepackage{xspace}
\usepackage{multicol}
\usepackage[compact]{titlesec}

\graphicspath{{figures/}}

\newcommand{\bi}{\begin{itemize}}
\newcommand{\ei}{\end{itemize}}
\newcommand{\eg}{e.g.\ }
\newcommand{\ie}{i.e.\ }

\titlespacing{\section}{2pt}{*0}{*0}

\begin{document}

\date{}

\title{\Large \bf EUI-64 Considered Harmful}

\author{
{\rm Erik C.\ Rye}\\
CMAND
\and
{\rm Jeremy Martin}\\
MITRE
\and
{\rm Robert Beverly}\\
Naval Postgraduate School
} 

\maketitle

\thispagestyle{empty}

\subsection*{Abstract}
This position paper considers the privacy and security
implications of EUI-64-based IPv6 addresses.  By
encoding MAC addresses, EUI-64 addresses
violate layers by exposing hardware identifiers in IPv6 addresses.
The
hypothetical threat of EUI-64 addresses is well-known,
and 
the adoption of privacy extensions in operating systems  (OSes)
suggests this vulnerability has been mitigated.
Instead, our work seeks to quantify the empirical existence of EUI-64
IPv6 addresses in today's Internet.  By analyzing: i) traceroutes; ii)
DNS records; and iii) mobile phone behaviors, we find surprisingly
significant use
of EUI-64.  We characterize the origins and behaviors of these EUI-64
IPv6 addresses, and advocate for changes in provider IPv6 addressing
policies.

\section{Introduction}

The privacy and security implications of Layer-2 hardware network identifiers,
\eg 48-bit IEEE Media Access Control (MAC) addresses~\cite{7297898}, have been
well-studied.  In particular, because these addresses are static, difficult for
the average user to change, and persist across network attachment points, they
can be exploited to both fingerprint devices (\eg identify the manufacturer and
model~\cite{acsac16furious}, thereby permitting targeted attacks) as well as
compromise user privacy by providing a persistent tracking identifier.  
\ifx\foo
The
potential vulnerabilities of MAC addresses are critical to understand as they
are used in all current popular link-layer protocols (\ie Ethernet, WiFi, Bluetooth).
For instance, recent work has shown the ability to: correlate cellular and MAC
identifiers \cite{gsm80211correlation-milcom13, o2017mobile,
rajavelsamy2018privacy, baek2018wi}, correlate virtual access points
\cite{roth2017graph}, and track mobile devices \cite{martin2017study, mims_2012} by
leveraging MAC addresses.  In response, vendors have responded by randomizing
MAC addresses and their use~\cite{gruteser2005enhancing}.  Despite these
efforts, security researchers have found vulnerabilities in the randomization
itself~\cite{vanhoefRandomMAC, martin2017study}.

While MAC addresses are used at the link-layer, and therefore should only be
visible to entities able to snoop on the local network, the global uniqueness
property of MAC addresses has inspired their use beyond the link-layer.  One
such example, widely documented by researchers is the usage of device
identifiers (MAC, IMEI, IMSI) by advertisement analytic platforms within mobile
applications~\cite{continella2017obfuscation, papadopouloslong}.  This practice
exposes the link-layer address to eavesdropping attacks external to the local
network.
\fi

This work focuses on MAC addresses in IPv6.  Auto-configured IPv6 addresses
were traditionally formed by encoding the interface's 48-bit MAC address as an
EUI-64~\cite{rfc7042} into the lower order bits of the 128-bit IPv6 address as
part of Stateless Address Autoconfiguration (SLAAC)~\cite{rfc4862}.  Today,
however, modern OSes randomize the host bits of their IPv6 addresses according
to the privacy extensions standard~\cite{rfc4941}. These addresses are
randomized and highly ephemeral~\cite{Plonka:2015:TSC:2815675.2815678}.

Most recently, two works at the ACM IMC 2018 discover non-trivial numbers of
EUI-64 addresses within large-scale IPv6 topology discovery
campaigns~\cite{imc18beholder,gasser2018clusters}.  This unexpected result is
due to the fact that while end-hosts have largely abandoned the use of EUI-64
addresses, Customer Premise Equipment (CPE) infrastructure (\eg home routers
and gateways) still make extensive use of such addresses.  Critically, the MAC
addresses of these devices is revealed remotely when they respond to traceroute
probing.

While revealing these addresses is seemingly innocuous at first blush, they
introduce two potential vulnerabilities: i) the ability to identify the
manufacturer and model of a device, thereby permitting targeted attacks; and
ii) the ability to track users despite efforts to prevent such tracking.  More
specifically, not only are the host bits of a client's IPv6 address ephemeral,
their allocated network prefix is similarly
ephemeral~\cite{Padmanabhan:2016:RDA:2987443.2987461}.  Indeed, many providers
induce their client's IPv4 and IPv6 addresses to change regularly for privacy
reasons~\cite{zwangstrennung}.  However, while the client's end-host IPv6
address is changing, the address of her home gateway is not -- thus an attacker
able to perform traceroutes to the client can determine that it is the same
client, and track the client's assigned addresses over time.

In this preliminary survey, we study the use, prevalence, and potential
vulnerabilities of EUI-64 addresses among four separate sources of data:
\bi
\addtolength{\itemsep}{-0.5\baselineskip}
  \item CAIDA Archipelago traceroute data from December 2008 to July 2018
  \item Traceroutes to RIPE nodes conducted in 2018
  \item ANY and AAAA RR lookups from Rapid7's fDNS data set between October 2016
    and December 2018
  \item Inspection of 27 mobile devices in a lab environment
\ei

\section{Infrastructure EUI-64 Prevalence}
\label{sec:prevalence}

Plonka and Berger report that fewer than 1\% of clients accessing a large CDN
in 2015 via IPv6 use addresses containing EUI-64
MACs~\cite{Plonka:2015:TSC:2815675.2815678}.  While the population of clients
with EUI-64 addresses is small, recent work found non-trivial numbers within
IPv6 \emph{infrastructure} \cite{imc18beholder,gasser2018clusters}.  We first
analyze the macro composition of two large public datasets.

The CAIDA traceroute data \cite{caida-topo} contains 3,650,488 unique MAC
addresses appearing in 5,794,444 distinct EUI-64 IPv6 addresses between
December 2008 and July 2018.  Between the months of October 2016 and December
2018, Rapid7's fDNS dataset \cite{rapid7} contains 50,060,121 AAAA records with
an EUI-64 address, with 7,705,481 unique EUI-64 IPv6 addresses containing
2,177,516 distinct MAC addresses. 



The distribution of EUI-64 IPv6 addresses is non-uniform; some autonomous
systems (AS) are home to far greater numbers of active EUI-64 addresses than
others.  Over 2.8 million EUI-64 addresses appear in the CAIDA data from
one provider's prefixes, whereas only dozens appear in other networks.
This appears to be indicative of the manufacturer of the prominent CPE devices
issued by the respective service provider.  The most common providers and
manufacturers discovered in the CAIDA data are highlighted below: 

\begin{tabular}{l|l}
  \textbf{Providers} \footnotemark & \textbf{Manufacturers} \\\hline
  AS A (50\%) & Actiontec (57\%)      \\
  AS B (11\%) & AVM (FRITZ!) (14\%)     \\
  AS C (9\%)  & Arcadyan (8\%)     \\
  AS D (6\%) & Huawei (6\%)\\
  AS E (5\%) & ZTE (3\%)\\
\end{tabular}
\footnotetext{ASNs anonymized}


\section{Tracking users over CPE Address Changes}
To highlight tracking and privacy implications of CPE-based EUI-64 usage we
conduct a longitudinal IPv6 traceroutes to fixed nodes.  We fetch the
set of RIPE Atlas nodes~\cite{ripeatlas} with working IPv6 daily, recording
their address and Atlas node identifier.  The node identifier allows us to
continually probe the same end node over time, even if its address changes.  We
traceroute to each node and find the penultimate address when the traceroute
reaches the destination.  In this fashion, we can determine when service
providers change the address of the link between the provider and the customer's
CPE.  We are particularly interested in cases where the service provider changes
the assigned CPE prefix, but the CPE's address is EUI-64.

These prefix changes, assumedly meant to provide privacy protections, are
trivially enumerated due to the CPE device responding to network probing using
an EUI-64 derived address.  Furthermore, and of particular concern, this
behavior allows for the potential to track client devices, even those using
ephemeral addresses, as an implicit association to the CPE's EUI-64 address can
be inferred via the derived network path.  Depicted below, traces to the same
unique host over the course of a month, illustrate a representative behavior in
which the CPE device consistently flips between two /48 but distinct /64 IPv6
network prefixes on a daily basis. 

\begin{tabular}{l | l}
  \textbf{Date} & \textbf{CPE IPv6 /64 Prefix}\footnotemark \\\hline
  20181024 & \texttt{xxxx:xxxx:xxx1:1676} \\
  20181025 & \texttt{xxxx:xxxx:xxx2:46e1}  \\
  \ldots   & \dots \\
  20181102 & \texttt{xxxx:xxxx:xxx1:a6d7} \\
  20181103 & \texttt{xxxx:xxxx:xxx2:c9b9} \\
\end{tabular}
\footnotetext{IPv6 prefixes anonymized}

\noindent Other providers assign new /64s within the same /48 every day, while
we even observe CPEs being assigned prefixes in different /16s owned
by their provider.


Conversely, we detect single MAC addresses appearing in EUI-64 IPs
spanning multiple ASNs and countries in traceroutes occurring within a short
time interval. In this case, it is unlikely that there is a single device
traveling to disparate locations.  Rather, we posit that these devices auto
configure to a default hardware address, or represent devices that have
inappropriately reused a MAC address.

\section{Mobile Phone Behaviors}
Support for privacy extensions has been available since iOS 4.3 and Android
4.0~\cite{internet_society}, suggesting that privacy concerns have long been
addressed on mobile devices. While all 28 mobile device models, produced by 12
manufacturers, that we tested implemented privacy extensions for the primary
IPv6 address, only the OnePlus 6 running Android 8.1 had \emph{only} privacy
extension addresses.  All 27 other models, covering Android 4.3 to 9.0, had
secondary, global EUI-64 addresses assigned to their interfaces. Interestingly,
this EUI-64 address was not always visible from the system settings application;
instead, we connected to each phone using the Android Debugger and used
the IPv6 addresses discovered via the debugger as ground truth.





After obtaining each device's EUI-64 IPv6 address (except the OnePlus 6), we
successfully elicited an ICMPv6 Echo Response from each of the devices in
response to an Echo Request addressed to its EUI-64 IPv6 address while connected
to an 802.11 wireless network. Early experiments involving Google Pixel phones
connected to LTE networks suggest the ability to \texttt{ping6} EUI-64
IP addresses on mobile devices on cellular networks, as well. This implies a
potential to track individual devices, and hence users, as they move across
networks.  For example, an adversary who knows the MAC address of her target
might deduce a range of network prefixes the target might be in; exhaustive
probing of these prefixes with the lower bits set to the EUI-64 generated by the
target's MAC address quickly confirms or rules out that device's presence.



\section{Future Work}
A more robust quantification of service provider-induced network prefix changes
utilizing our EUI-64 CPE address tracking mechanism remains in progress.
Lastly, we hope to investigate the plausibility of the EUI-64 targeted
attack described above, in which \emph{a priori} knowledge of mobile device's
MAC address is assumed.   

\section{Disclaimer}
Views and conclusions are those of the authors and should not be interpreted as
representing the official policies or position of the U.S. Government. The
author's affiliation with The MITRE Corporation is provided for identification
purposes only, and is not intended to convey or imply MITRE's concurrence with,
or support for, the positions, opinions or viewpoints expressed by the author.

{\footnotesize \bibliographystyle{acm}
\bibliography{furiousmac}}

\end{document}